\documentclass[prd,showpacs,showkeys,floatfix,nofootinbib,eqsecnum,
               a4,preprint,12pt,tightenlines,fleqn]{revtex4-1}
\usepackage{amssymb,epsf}
\usepackage{latexsym,,amssymb}
\usepackage{amsmath,mathrsfs}
\usepackage{graphics,epsfig}
\usepackage[english]{babel}

\usepackage{slashed}
\usepackage{mathtools}

\DeclarePairedDelimiter\floor{\lfloor}{\rfloor}

\newcommand{\version}{v6}  

\newcommand{\beq}{\begin{equation}}
\newcommand{\eeq}{\end{equation}}
\newcommand{\beqa}{\begin{eqnarray}}
\newcommand{\eeqa}{\end{eqnarray}}
\newcommand{\bsubeqs}{\begin{subequations}}
\newcommand{\esubeqs}{\end{subequations}}

\begin{document}
\noindent  Phys. Rev. D 91, 045028 (2015)
\hfill arXiv:1412.1008\;(\version)\newline\vspace*{2mm}
%
%
\title{\vspace*{2mm}
Fermions with a bounded and discrete mass spectrum
\\[2mm]
\vspace*{2mm}}

\author{F.R.~Klinkhamer}
\email{frans.klinkhamer@kit.edu}
\author{L. Yang}
\email{lu.yang@kit.edu}
\affiliation{Institute for
Theoretical Physics, Karlsruhe Institute of
Technology (KIT), 76128 Karlsruhe, Germany\\}

\begin{abstract}%
\noindent
\vspace*{-4mm}\newline
A mechanism for determining fermion masses
in four spacetime dimensions is presented, which uses a
scalar-field domain wall extending in a fifth spacelike dimension
and a special choice of Yukawa coupling constants.
A bounded and discrete fermion mass spectrum is obtained analytically for spinors localized in the fifth dimension.
These particular mass values depend on a combination of
the absolute value of the Yukawa coupling constant
and the parameters of the scalar potential.
A similar mechanism for a finite mass spectrum
may apply to $(1+1)$--dimensional fermions
relevant to condensed matter physics.
\end{abstract}

\pacs{11.27.+d, 12.15.Ff, 11.30.Rd, 68.37.-d}   


\keywords{domain wall, fermion mass spectrum, chiral fermions, condensed matter physics}

\maketitle

\section{Introduction}
\label{sec:Introduction}

There are at least two ways to determine fermion masses explicitly:
Dirac's quantization of the spinor mass over a
four-dimensional space embedded in a five-dimensional  
projective space~\cite{Dirac1936} and the Kaluza--Klein model with a fifth periodic compact dimension
(see, e.g., Ref.~\cite{Rubakov2001} and references therein).
Both of these fermion mass spectra are discrete but unbounded
($|m_n| \to \infty$ for $n \to \infty$).

In this article, we consider another possibility by adding
one extra spacelike \emph{open} dimension to the standard Minkowski spacetime and introducing Yukawa couplings between the scalar and the Dirac fermions.
Our suggestion relies on having a scalar-field domain-wall background in the five-dimensional spacetime and making a special choice for the values of the two Yukawa coupling constants (they must be the opposite of each other).
A similar five-dimensional setup has, of course, been used in previous studies~\cite{Rubakov2001,Kaplan1992}.
Restricting to \emph{localized} fermion solutions   
in the background of the scalar-field domain wall,
we find a discrete and bounded fermion mass spectrum.

These three ways of calculating fermion masses
have one thing in common: the dimension of spacetime
is increased from four to five or more.
Note that
this fifth extra dimension may be essentially different
from the four dimensions of standard Minkowski spacetime.
In more general terms, an infinite number of degrees
of freedom has been added to the four-dimensional theory,
even though the theory considered in the end
applies again to four spacetime dimensions.

The outline of our short paper is as follows.
In Sec.~\ref{sec:Notation}, we establish our notation.
In Sec.~\ref{sec:Mechanism}, we define the theory and look for
nonsingular localized fermion
solutions in the background of a scalar-field domain wall.
The main results of this section are the bounded
fermion mass spectra \eqref{eq:integer-mass-F-N}
and \eqref{eq:mass-spectrum-noninteger-degree},
together with the corresponding wave functions \eqref{eq:vl-vr-solutions}
and  \eqref{eq:vl-vr-solutions-noninteger-degree}.
A further result of this section is the vanishing backreaction  
\eqref{eq:fermion-source-term-classical} of the fermions on the scalar-field domain wall,
at least as far as
the classical dynamical equations are concerned.
In Sec.~\ref{sec:Discussion}, we present some further remarks
and comment on the possible relevance to condensed matter physics.
Appendix~\ref{app:Integrable-solutions-with-noninteger-Legendre-degree}
contains the details of the domain-wall fermion solutions
for a generic ratio of the Yukawa and Higgs coupling constants.
Appendix~\ref{app:Massless-spinor-solution} recalls, for completeness, the massless chiral spinor solution~\cite{JackiwRebbi1976}.

Before we start, it may be worthwhile to
emphasize that our main interest is in finding a mechanism
responsible for a bounded and discrete fermion mass spectrum   
(for earlier results in the brane-world context, see, e.g.,
Ref.~\cite{RingevalPeterUzan2001}).
Even though our elementary discussion
is far removed from reality (fermions of a chiral
gauge theory are not considered)
and heuristic to a certain extent (leaving open
the proper role and interpretation of the additional
degrees of freedom, naively those from an extra dimension),
we believe that there is some interest in finding quite
generally that a bounded and discrete fermion mass spectrum   
may result from a simple normalization condition.

\section{Notation}
\label{sec:Notation}

We take $w$ to denote the extra spacelike coordinate and keep
$x^\mu$ for the coordinates of the usual four-dimensional Minkowski spacetime.
Latin indices $a,b,\ldots$ refer to all five spacetime coordinates,
while Greek indices $\mu,\nu,\ldots$ leave out the fifth coordinate $w$.
With the spacetime coordinates
\bsubeqs\label{eq:5Dcoordinates-metric}
\beqa
\label{eq:5Dcoordinates}
(x^a) &=& (x^{0},x^{1},x^{2},x^{3},w)\,,
\eeqa
the metric of the flat five-dimensional spacetime $\mathcal{M}_5$ is given by
\beqa
\label{eq:metric}
 \left( g_{a b}\right) &=& \text{diag}(1,-1,-1,-1, -1) \,.
\eeqa
\esubeqs

Next, we specify the $2\times 2$ Pauli  and the $4\times 4$ Dirac
matrices to be used:
\bsubeqs\label{eq:Pauli-Dirac-matrices}
\beqa
\label{eq:Pauli-matrices}
\sigma^1 &=&\left(\begin{array}{cc} 0 & 1\\ 1 & 0\end{array} \right) , \ \ \sigma^2=\left(\begin{array}{cc}0 & -i\\ i & 0\end{array}\right), \ \ \sigma^3=\left(\begin{array}{cc} 1& 0\\ 0 & -1 \end{array} \right)
 \,,\\[2mm]
 \label{eq:Dirac-matrices}
\gamma^{0}  &=&
    \left(\begin{array}{cc}0 & \mathbb{I}\\ \mathbb{I} & 0 \end{array} \right) , \ \
    \gamma^i=
    \left(\begin{array}{cc}0 & \sigma^i\\ -\sigma^i & 0\end{array} \right), \ \ \gamma^5=
    \left(\begin{array}{cc}-\mathbb{I}& 0 \\ 0 &\mathbb{I} \end{array} \right) \,,
\eeqa
\esubeqs
where $\mathbb{I}$ is the rank-two  identity matrix.
The five-dimensional gamma matrices corresponding to the metric \eqref{eq:metric} are then
\begin{equation}\label{eq:Gammas}
\Gamma^{\mu}= \gamma^{\mu} \,,\quad
\Gamma^{5}= -i\gamma^{5} \,.
\end{equation}
The factor $i$ in $\Gamma^{5}$ is to keep up with our choice of metric and the minus sign has been chosen for the sake of convenience.

Throughout, natural units are used with $\hbar=c=1$.

\section{Mechanism}
\label{sec:Mechanism}

\subsection{Theory}
\label{subsec:Theory}

Our starting point is the following
classical
Lagrange density over the
five-dimensional spacetime $\mathcal{M}_5$:
\begin{equation}\label{eq:L}
\mathcal{L}_5 =
\bar{\Psi}\, i\slashed{\partial}\,  \Psi
+\bar{\Omega}\, i\slashed{\partial}\,  \Omega
- f\, \bar{\Psi}\Psi\, \phi + f\, \bar{\Omega}\Omega\, \phi + \frac{1}{2}\,\partial_{a}\phi \, \partial^{a}\phi
-\frac{\lambda^2}{2}\, \left(\phi^2-M^2\right)^2 \,.
\end{equation}
This Lagrange density describes two spinor fields
$\Psi$ and $\Omega$
coupled to a real Higgs-like scalar field $\phi$
(the Yukawa coupling constant $f$ of the $\Omega$ scalar is taken
positive, $f > 0$). All fields are
defined over a five-dimensional flat spacetime.
The slashed differential operator in \eqref{eq:L} is explicitly $\slashed{\partial} \equiv \Gamma^{a}\,\partial_a$, with the
Dirac matrices $\Gamma^{a}$ as given by \eqref{eq:Gammas}.

As is clear from the preceding paragraph, we consider the
fields $\phi$, $\Psi$, and $\Omega$ to be \mbox{c-number} functions.
The relevance of our results to quantum field theory
will be briefly discussed in Sec.~\ref{subsec:Remarks-on-the-quantum-theory}.

It is also possible to add gauge fields or even a
dynamical metric field,
also considered to be \mbox{c-number} functions.
The simplest possibility is to add
a $U(1)$ gauge field and to have opposite electric charges
for the two spinors $\Psi$ and $\Omega$
and zero electric charge for the scalar $\phi$.
The present paper, however, keeps the types of fields to a minimum:
scalar and spinor.

The two spinor fields $\Psi$ and $\Omega$ are introduced with
strictly opposite Yukawa coupling constants in \eqref{eq:L},
for the sake of obtaining exact
classical
solutions later on (specifically, the two spinor-field source terms
in the scalar field equation  will cancel).
The positive parameters $M$, $f$, and $\lambda$ in \eqref{eq:L}
are the vacuum expectation value of the scalar field
and the coupling constants in the usual four-dimensional sense.
In the five-dimensional theory, these parameters acquire different mass dimensions compared to their counterparts in the four-dimensional theory. For instance, $\lambda$
then has mass dimension $-1/2$ and $M$ has mass dimension
$3/2$. Thus, $\lambda M$ may serve as a mass parameter.

The quantum theory corresponding to \eqref{eq:L}
is, most likely, nonrenormalizable,
given the presence of negative-mass-dimension couplings
such as $\lambda^2$ for the quartic scalar interaction term.
This is perhaps one more reason to consider the
fifth dimension to be essentially different than the
four of standard Minkowski spacetime, but this
suggestion will not be pursued further in this paper.

\subsection{Scalar-field domain-wall solution}
\label{subsec:Scalar-field-domain-wall-solution}

From the Lagrange density \eqref{eq:L},
the classical scalar field equation  is
\begin{equation}\label{eq:phi-eq}
-\partial_{a}\partial^{a} \phi
-2\,\lambda^2\, \left(\phi^2 -M^2\right)\, \phi
=
f\, \left(\bar{\Psi}\Psi-\bar{\Omega}\Omega\right) \,.
\end{equation}
As will be demonstrated in Sec.~\ref{subsec:Consistency-of-scalar-and-spinor-solutions}, the contributions of the
\mbox{c-number}
fields $\Psi$ and $\Omega$
 on the right-hand side of the scalar field equation  \eqref{eq:phi-eq}
can be made to cancel each other, so we drop them for the moment.
Assuming $\phi$ to depend only on the fifth coordinate $w$, Eq.~\eqref{eq:phi-eq} then becomes
\begin{equation}
\partial^{2}_w\phi(w)=2\lambda^2\;\left[\phi(w)^2-M^2\right]\;\phi(w) \,,
\end{equation}
with a domain-wall solution
\begin{subequations}\label{eq:domain-wall}
\begin{eqnarray}
\phi(w)&=&M\tanh(c\, w)\,,\\[2mm]
c   &\equiv&\lambda\, M\,,
\end{eqnarray}
\end{subequations}
where the constant $c$ has mass dimension $1$.
In order to have this domain-wall solution, it is necessary to choose the fifth dimension to be spacelike
(see, e.g., Ref.~\cite{Rubakov2001} for further discussion).

In the following subsections, we
focus on localized normalizable spinor
solutions in the background of a scalar-field domain wall.
There are, of course, also scattering-type
solutions~\cite{Rubakov2001},
but they have higher energies and reach out to $|w| = \infty$.

\subsection{Spinor \textit{Ans\"{a}tze}}
\label{subsec:Spinor-Ansaetze}

In the background of the scalar domain-wall solution \eqref{eq:domain-wall}, we have for the spinors
\begin{subequations}\label{eq:PsiOmega-eq}
\begin{eqnarray}
\left (i\gamma^{\mu}\partial_{\mu}+i\Gamma^5\partial_w\right)\,  \Psi &=&+f\, M\tanh(c\, w)\,  \Psi \label{eq:Psi-eq}\,,
 \\[2mm]
\left(i\gamma^{\mu}\partial_{\mu}+i\Gamma^5\partial_w\right)\,  \Omega &=&-f\, M\tanh(c\, w)\,  \Omega \,.
 \label{eq:Omega-eq}
\end{eqnarray}
\end{subequations}
From these last two equations,
it is clear that $f\, M$ has mass dimension $1$.

If we now write $\Psi$ and $\Omega$ in terms of
two-component spinors,
\beq\label{eq:Psi-Omega}
\Psi= \left( \begin{array}{c} \psi_{l}  \\  \psi_{r} \end{array} \right)
\,, \quad
\Omega= \left( \begin{array}{c} \omega_{l}  \\ \omega_{r} \end{array} \right) \,,
\eeq
Eq.~\eqref{eq:Psi-eq} for the $\Psi$ field  becomes
\bsubeqs\label{eq:psi-l-psi-r-equations}
\beqa
 i\sigma^{\mu}\partial_{\mu}\psi_{r}-\partial_{w}\psi_{l}&=&
 f\, M \tanh(c\, w)\,  \psi_{l} \,, \\[2mm]
 i\overline{\sigma}^{\mu}\partial_{\mu}\psi_{l}+\partial_{w}\psi_{r}&=&
 f\, M \tanh(c\, w)\,  \psi_{r} \,,
\eeqa
\esubeqs
where the index $\mu$ runs over $0,1,2,3$ (corresponding to the four coordinates of the usual Minkowski spacetime) and
\bsubeqs\label{eq:sigma-mu}
\beqa
 \sigma^{\mu} &\equiv &
(\mathbb{I}, \sigma^1, \sigma^2,\sigma^3)\,,
\\[2mm]
  \overline{\sigma}^{\mu}&\equiv& (\mathbb{I}, -\sigma^1, -\sigma^2,-\sigma^3)\,.
\eeqa
\esubeqs
Equation~\eqref{eq:Omega-eq} for the $\Omega$ field
acquires a similar form as \eqref{eq:psi-l-psi-r-equations},
with $(\psi_{l} \,,\, \psi_{r})$
replaced by $(\omega_{l} \,,\, \omega_{r})$ and $f$ by $-f$.

The next step is to make the following \textit{Ansatz}
(separation of variables):
\bsubeqs \label{eq:Psi-5D-separation-Ansatz}
\beqa
\label{eq:Psi-5D-separation-Ansatz-l}
\psi_{l}(x^{0},x^{1},x^{2},x^{3},w)&=&
v_{l}(w)\,\,  \chi(x^{0},x^{1},x^{2},x^{3}) \,,
\\[2mm]
\label{eq:Psi-5D-separation-Ansatz-r}
\psi_{r}(x^{0},x^{1},x^{2},x^{3},w)&=&
v_{r}(w)\,\,  \xi(x^{0},x^{1},x^{2},x^{3}) \,.
\eeqa
\esubeqs
For later use, we already give the corresponding
\textit{Ansatz} for the two-spinors of the $\Omega$ field:
\bsubeqs \label{eq:Omega-5D-separation-Ansatz}
\beqa
\label{eq:Omega-5D-separation-Ansatz-l}
\omega_{l}(x^{0},x^{1},x^{2},x^{3},w)&=&v_{l}'(w)\, \chi'(x^{0},x^{1},x^{2},x^{3}) \,,
\\[2mm]
\label{eq:Omega-5D-separation-Ansatz-r}
\omega_{r}(x^{0},x^{1},x^{2},x^{3},w)&=&v_{r}'(w)\, \xi'(x^{0},x^{1},x^{2},x^{3}) \,.
\eeqa
\esubeqs
Continuing with the discussion of the $\Psi$ field and
substituting the \textit{Ansatz} \eqref{eq:Psi-5D-separation-Ansatz}
into Eq.~\eqref{eq:psi-l-psi-r-equations}, we obtain:
\bsubeqs\label{eq:vchixi}
\beqa
 i\sigma^{\mu}\partial_{\mu}\xi(x)\,  v_{r}(w)
 -\partial_{w}v_{l}(w)\, \chi(x)
 &=&f\, M \tanh(c\, w) \, v_{l}(w)\, \chi(x) \,, \\[2mm]
 i\overline{\sigma}^{\mu}\partial_{\mu}\chi(x)\, v_{l}(w)
 +\partial_{w}v_{r}(w)\, \xi(x)&=&f\, M \tanh(c\, w)\,  v_{r}(w)\, \xi(x) \,,
\eeqa
\esubeqs
with function argument $(x)$ standing for $(x^{0},x^{1},x^{2},x^{3})$.

We make one more \textit{Ansatz} at this point:
\bsubeqs\label{eq:4D-Ansatz}
\beqa
\label{eq:4D-Ansatz-partial-xi}
 i\sigma^{\mu}\partial_{\mu}\xi(x) &=& m_{4}\,  \chi(x)  \,,
 \\[2mm]
\label{eq:4D-Ansatz-partial-chi}
 i\overline{\sigma}^{\mu}\partial_{\mu}\chi(x) &=& m_{4}\, \xi(x)\,,
\eeqa
together with the standard normalization condition
on the two-spinors   
\beqa
\label{eq:normalization-condition}
\chi^\dagger(x)\, \xi(x)+\xi^\dagger(x)\, \chi(x) \ &=& 2\, |m_{4}|\,.
\eeqa
\esubeqs
We could have assigned two different parameters at the places of $m_{4}$ in
Eqs.~\eqref{eq:4D-Ansatz-partial-xi} and \eqref{eq:4D-Ansatz-partial-chi}.
However, if we wish to interpret $\chi$ and $\xi$ as the left-handed  and right-handed components of the same massive Dirac spinor in four-dimensional Minkowski spacetime,
it is necessary to have a unique mass value $m_{4}\ne 0$
in Eqs.~\eqref{eq:4D-Ansatz-partial-xi} and \eqref{eq:4D-Ansatz-partial-chi}.
Equation~\eqref{eq:vchixi} is now reduced to the following set of
coupled equations:%
\bsubeqs\label{eq:firstordereqforv}
 \beqa
 -\partial_w v_{l}(w) + m_{4}\, v_{r}(w) &=& f\, M\tanh(c\, w) \,  v_{l}(w) \,,\\[2mm]
  +\partial_w v_{r}(w) + m_{4}\, v_{l}(w)  &=& f\, M\tanh(c\, w) \,  v_{r}(w)  \,.
\eeqa\esubeqs

In order to solve this last set of equations,
we make a change of variable
\beq \label{eq:newvariable}
 s\equiv\tanh(c\, w)  \,,
\eeq
so that $\partial_w = c\,(1-s^2)\partial_s$.
The equations then take a more recognizable form:
\bsubeqs\label{eq:partial-vl-vr-eqs}
\beqa
\label{eq:partial-vl-eq}
-(1-s^2)\, \partial_s v_{l}(s) + m_f\,  v_{r}(s)&=& F \, s \,  v_{l}(s) \,,
 \\[2mm]
\label{eq:partial-vr-eq}
+(1-s^2)\, \partial_s v_{r}(s) + m_f\,  v_{l}(s)&=& F \, s \,  v_{r}(s)\,,
\eeqa\esubeqs
where the new parameters are
\bsubeqs\label{eq:mf-M-definitions}
\beqa
\label{eq:mf-definition}
m_f &\equiv& m_{4}/c \equiv m_{4}/(\lambda M)\,,
\\[2mm]
\label{eq:M-definition}
F   &\equiv&f/\lambda\,.
\eeqa\esubeqs
Note that both $m_f$ and $F$ are dimensionless.

From  the first equation \eqref{eq:partial-vl-eq}
we can express $v_{r}$ in terms of $v_{l}$ and $\partial_s v_{l}$. Then, we substitute this expression for $v_{r}$
into the second equation \eqref{eq:partial-vr-eq} and obtain
\begin{equation} \label{eq:vl-Legendre-equation}
(1-s^2)\, \frac{d^2 v_{l}(s)}{d s^2} -2 s\,  \frac{dv_{l}(s)}{d s}
+\left[F(F+1) - \frac{F^2-m_{f}^2}{1-s^2}\right]\,v_{l}(s) =0 \,.
\end{equation}
A similar equation for $v_{r}$ is obtained as:
\begin{equation}\label{eq:vr-Legendre-equation}
(1-s^2)\, \frac{d^2 v_{r}(s)}{d s^2} -2 s\,  \frac{dv_{r}(s)}{d s}
+\left[F(F-1) - \frac{F^2-m_{f}^{2}}{1-s^2}\right]\,v_{r}(s) =0 \,.
\end{equation}
The above two uncoupled equations will be solved
in the next subsection and Appendix~\ref{app:Integrable-solutions-with-noninteger-Legendre-degree}.

\subsection{Spinor solutions and fermion mass spectrum}
\label{subsec:Spinor-solutions-and-mass-fermion-spectrum}

The two equations \eqref{eq:vl-Legendre-equation} and \eqref{eq:vr-Legendre-equation} belong to the class of Legendre equations~\cite{Erdelyi1953,AbramowitzStegun1965,MagnusOberhettingerSoni1966},
with degree $\nu$ and order $\mu$ taking the values
\bsubeqs\label{eq:nu-mu}
\beqa
\nu &\in& \{F, F-1 \}\,,
\\[2mm]
\mu &=& \pm\,\sqrt{F^{2}-m_{f}^{2}} \,.
\eeqa
\esubeqs
These equations admit square-integrable solutions on the open interval
$(-1,\,1)$ with the measure induced by the reparametrization
\eqref{eq:newvariable} if and only if the absolute values of the degree
$\nu$ and order $\mu$ differ by integers
and the degree is bounded from below, $\nu > 1/2$.  
For simplicity, we focus on the case of integer degree in this
subsection. The case of noninteger degree is discussed
in Appendix~\ref{app:Integrable-solutions-with-noninteger-Legendre-degree},
which also gives the standard form \eqref{eq:GL}
of the Legendre equation and the corresponding definitions of
the degree $\nu$ and the order $\mu$.

Assuming the degree to be integer and restricting to square-integrable solutions, the possible absolute values for the degree and order are
given by (cf. Table~4.8.2 in Ref.~\cite{MagnusOberhettingerSoni1966})
\bsubeqs\label{eq:integers}
\beqa
F &=& l \in \mathbb{N}_0 \backslash \{0,\, 1 \} \,,\\[2mm]\
\sqrt{F^{2}-m_{f}^{2}}   &=& m \,,\\[2mm]
m &\in&  \{ 1, 2, \ldots l-2, l-1 \}  \,,
\eeqa
\esubeqs
where some of the usual integers have been omitted
(specifically, $l=0,1$ and $m=0,l$).

The reason for omitting certain integers in \eqref{eq:integers}
is twofold: first, there are
\emph{two} Legendre equations to consider simultaneously
[namely Eqs.~\eqref{eq:vl-Legendre-equation} and \eqref{eq:vr-Legendre-equation}]
and, second, the relevant solutions [the associated Legendre
functions of the first kind, $P_{l}^m(s)$] are required to
have a finite norm for an $s$--measure equal to $dw/ds=1/(1-s^2)$.
Specifically, we have the following normalization of the relevant solutions:
\beq\label{eq:Plm-norm}
\int_{-1}^{1}\frac{d s}{(1-s^2)}\,\Big(P_{l}^m(s)\Big)^2
=\frac{(l+m)!}{m\, (l-m)!} \;,
\eeq
for positive integers $l$ and $m$.  
In particular, the degree $m$ of the solution $P_{l}^m(s)$
cannot be $0$; otherwise, the norm of the corresponding Legendre
function would be infinite for the $s$--measure  
inherited from the five-dimensional spacetime.

According to Eqs.~\eqref{eq:4D-Ansatz-partial-xi}
and \eqref{eq:4D-Ansatz-partial-chi},
we interpret $m_{4}$ entering the $m_f$ definition \eqref{eq:mf-definition}
as the inertial mass of a fundamental fermion propagating
in the usual four-dimensional Minkowski spacetime.
The above conditions \eqref{eq:integers}  then imply that,
since $F\equiv f/\lambda$ is a fixed integer $N$, there is only a finite number of fundamental fermions. The  masses $m_4$ of these fundamental fermions take the following values:
\bsubeqs\label{eq:integer-mass-F-N}
\begin{eqnarray}
\label{eq:integer-mass}   
\hspace*{-5mm}
m_4/(\lambda M) &\in&
\left\{
\pm \sqrt{N^2-1^2}, \pm\sqrt{N^2-2^2}, \pm
\sqrt{N^2-3^2},\, ...  \,,\,\pm\sqrt{N^2-(N-1)^2}
\right\} \,,
\\[2mm]
\label{eq:integer-F}
\hspace*{-5mm}
f/\lambda  &=&  N  \,,
\\[2mm]
\label{eq:integer-N}
\hspace*{-5mm}
N   &\in& \left\{ 2,3,4,\ldots \right\}\,.
\end{eqnarray}
\esubeqs
The corresponding solutions of Eqs.~\eqref{eq:vl-Legendre-equation} and \eqref{eq:vr-Legendre-equation} are:
\bsubeqs\label{eq:vl-vr-solutions}
\beqa
\label{eq:vl-solution}
v_{l}(s) &\propto& P_{F}^{\,\pm\,\sqrt{F^2-m_{f}^2}}(s)\,,
\\[2mm]
\label{eq:vr-solution}
v_{r}(s) &\propto& P_{F-1}^{\,\pm\,\sqrt{F^2-m_{f}^{2}}}(s)\,,
\eeqa\esubeqs
with $s=s(w)$ as defined by Eq.~\eqref{eq:newvariable}
and $F \equiv f/\lambda$ and $m_f \equiv m_{4}/(\lambda M)$
taking values according to \eqref{eq:integer-mass-F-N}.  

It is interesting to note that the values $0$ and $f\, M$ for $m_4$
do not appear in the fermion mass spectrum \eqref{eq:integer-mass}
due to the normalizability condition on $v_{r}(w)$ and $v_{l}(w)$.
Since $v_{r}(s)$ is given by \eqref{eq:vr-solution},
$m_f$ cannot be $0$ [the function $P_{l}^m$ would have $|m|>l$].   
The same expression for $v_{r}(s)$ also
tells us that $m_f$ cannot be equal to $F$
[$P_{F-1}^{0}$ is not normalizable for the relevant measure, according to
Eq.~\eqref{eq:Plm-norm}].
Hence, both the minimal mass value ($m_4=0$)
and  the maximal mass value ($m_4=f M=F \lambda M$)  are not present in the mass spectrum of the localized Dirac spinors
(there does exist a massless spinor solution,
but it is chiral; see Appendix~\ref{app:Massless-spinor-solution}).
As to the range of $|m_4|/(\lambda M)$, the highest value is
$N\,\sqrt{1-1/N^2}$ and the lowest value is
$\sqrt{2N-1}$, so that the mass gap increases as
the ratio of coupling constants $f/\lambda = N$ grows.

With the wave functions $v_{l}(w)$ and $v_{r}(w)$
entering the chiral fields \eqref{eq:Psi-5D-separation-Ansatz-l} and \eqref{eq:Psi-5D-separation-Ansatz-r},
it is also possible to distinguish left-handed and right-handed fermions, because the corresponding solutions
\eqref{eq:vl-vr-solutions} are different.
This means that the left-handed and right-handed fermions are localized differently in the fifth dimension. This is not altogether surprising since the different chiralities trace back to the fact that left-handed and right-handed fermions correspond to different eigenvalues of the $\Gamma^5$ matrix, namely $-1$ and $+1$.
But $\Gamma^5$ is also the Dirac gamma matrix of the fifth dimension.
So, the double role of $\Gamma^5$ brings together chirality and the fifth dimension (see also Sec.~\ref{sec:Discussion} for further comments).
In the discussion up until now,
we have used one specific domain-wall solution \eqref{eq:domain-wall} with
a particular direction and this direction treats the left and right chiralities differently in a particular way. Using the other domain-wall solution obtained by $\phi \to -\phi$, would switch the roles of left and right chirality.

The square-integrable solutions for the case of noninteger degree $\nu$ are detailed in Appendix~\ref{app:Integrable-solutions-with-noninteger-Legendre-degree}.
Applying these solutions to
the two particular Legendre equations for $v_r(w)$ and $v_l(w)$ as
given by Eqs.~\eqref{eq:vl-Legendre-equation}
and \eqref{eq:vr-Legendre-equation},
the following fermion mass spectrum is obtained:
\beq
\label{eq:mass-spectrum-noninteger-degree}
m_4/(\lambda M)\in\left\{
\pm \sqrt{2F\times 1-1^2},\,
\pm \sqrt{2F\times 2-2^2},\,
...\,,\,
\pm \sqrt{2F\times \floor{F}-(\floor{F})^2} \right\} \,,
\eeq
where $F \equiv f/\lambda>3/2$ is a positive noninteger number  
and $\floor{F}$ denotes the largest integer that is
not greater than $F$.  The corresponding wave functions are%
\bsubeqs\label{eq:vl-vr-solutions-noninteger-degree}
\beqa
\label{eq:vl-solution-noninteger-degree}
v_{l}(s) &\propto& P_{F}^{\, - \,\sqrt{F^2-m_{f}^2}}(s)\,,
\\[2mm]
\label{eq:vr-solution-noninteger-degree}
v_{r}(s) &\propto& P_{F-1}^{\, - \,\sqrt{F^2-m_{f}^{2}}}(s)\,,
\eeqa\esubeqs
again for noninteger $F>3/2$ and with $m_f \equiv m_{4}/(\lambda M)$ from Eq.~\eqref{eq:mass-spectrum-noninteger-degree}.

Three remarks are in order.
First, the mass value $0$ again does not appear in the spectrum
\eqref{eq:mass-spectrum-noninteger-degree}.
The reason is that, for both spinor equations to have square-integrable  solutions, it is necessary that $\sqrt{F^2-(m_4/\lambda M)^2} \leq F-1.$
Second, the mass value $0$ does appear for a chiral
solution~\cite{JackiwRebbi1976}, whose basic structure is
recalled
in Appendix~\ref{app:Massless-spinor-solution}.
Third, taking the spectrum from Eq.~\eqref{eq:mass-spectrum-noninteger-degree}
as it stands and letting $F$ approach an integer $N\geq 2$ from below
reproduces precisely the masses from Eq.~\eqref{eq:integer-mass}.

\subsection{Consistency of scalar and spinor solutions}
\label{subsec:Consistency-of-scalar-and-spinor-solutions}

We now complete the discussion on the exactness and consistency of the classical solutions.
Let us perform the same calculation for the $\Omega$ field
as for the  $\Psi$ field. The $\Omega$ versions of Eqs.~\eqref{eq:psi-l-psi-r-equations} and \eqref{eq:vchixi} are obtained by replacing $f$ by $-f$.
Equally, replacing $F$ in Eqs.~\eqref{eq:vr-Legendre-equation} and \eqref{eq:vl-Legendre-equation} by $-F$ gives the equations for $\Omega$.
For a given $\Psi$ solution with functions
$v_{l}(w)$, $v_{r}(w)$, $\xi(x)$, and $\chi(x)$,
the corresponding $\Omega$ solution has the following primed wave functions:%
\bsubeqs\label{eq:primed-functions}
\beqa
\label{eq:vl-prime}
v'_{l}(w) &=&  v_{r}(w)\,,
\\[2mm]
\label{eq:vr-prime}
v'_{r}(w)  &=& v_{l}(w)\,,
\eeqa
\esubeqs
and similar relations for the primed two-spinor fields
as for the unprimed two-spinor fields:
\bsubeqs\label{eq:4D-Ansatz-prime}
\beqa
\label{eq:4D-Ansatz-partial-xi-prime}
 i\sigma^{\mu}\partial_{\mu}\xi'(x) &=& m_{4}\,  \chi'(x)  \,,
 \\[2mm]
\label{eq:4D-Ansatz-partial-chi-prime}
 i\overline{\sigma}^{\mu}\partial_{\mu}\chi'(x) &=& m_{4}\, \xi'(x)\,,
 \\[2mm]
\label{eq:normalization-condition-prime}
\chi'^{\,\dagger}(x)\,\xi'(x)+\xi'^{\,\dagger}(x)\, \chi'(x) &=& 2\, |m_{4}|\,,
\eeqa
\esubeqs
where the same mass value $m_{4}$ has been taken as for the
unprimed two-spinor fields in \eqref{eq:4D-Ansatz}.

Next, we expand the spinor-field source term of
the classical scalar field equation  \eqref{eq:phi-eq} and see that the two contributions cancel as follows:
\begin{eqnarray}\label{eq:fermion-source-term-classical}
f\,\left(\bar{\Psi}\Psi-\bar{\Omega}\Omega\right)&=& f\,\left[\psi_{l}^{\dagger}\psi_{r}+\psi_{r}^{\dagger}\psi_{l}
-\omega_{l}^{\dagger}\omega_{r}-\omega_{r}^{\dagger}\omega_{l}\right] \nonumber\\[2mm]
&=& f\,\left[v_{l} v_{r}\,  (\chi^{\dagger}\xi+\xi^{\dagger}\chi)
-v'_{l} v'_{r}\,  (\chi^{'\dagger} \xi^{'}    +\xi^{'\dagger}\chi^{'})\right]  \nonumber \\[2mm]
                                      &=& 0 \,,
\end{eqnarray}
where the second equality uses the fact that the wave functions
$v_{l,r}$  and $v'_{l,r}$ are real and the third equality follows from
\eqref{eq:primed-functions}, \eqref{eq:normalization-condition}, and \eqref{eq:normalization-condition-prime}.
Actually, even if we had flipped the sign of $m_{4}$
in Eqs.~\eqref{eq:4D-Ansatz-partial-xi-prime}
and \eqref{eq:4D-Ansatz-partial-chi-prime},
the third equality in \eqref{eq:fermion-source-term-classical} would still hold true.

In short, the scalar domain wall \eqref{eq:domain-wall}
and the obtained spinor fields \eqref{eq:Psi-Omega}
are exact solutions of the combined classical field equations
\eqref{eq:phi-eq} and \eqref{eq:PsiOmega-eq}.
These spinor fields have chiral components
\eqref{eq:Psi-5D-separation-Ansatz} and \eqref{eq:Omega-5D-separation-Ansatz}
with wave functions given by associated Legendre
functions \eqref{eq:vl-vr-solutions},
\eqref{eq:vl-vr-solutions-noninteger-degree}, and \eqref{eq:primed-functions}

\subsection{Remarks on the quantum theory}
\label{subsec:Remarks-on-the-quantum-theory}

The discussion of Sec.~\ref{sec:Mechanism} up to and including Sec.~\ref{subsec:Consistency-of-scalar-and-spinor-solutions}
has been for \mbox{c-number} fields
$\phi$, $\Psi$, and $\Omega$.
For the quantum theory, these fields become operators
and the discussion can follow that of  
the Jackiw--Rebbi paper~\cite{JackiwRebbi1976}. We now use the following states:
\beq
|P,\pm,\pm;\, p, m_{4};\,  k, \widetilde{m}_{4} \rangle\,,
\eeq
corresponding to a single soliton (domain wall) in five-dimensional
Minkowski spacetime with total four-momentum $P_{\mu}$   
and fourfold degeneracy~\cite{JackiwRebbi1976} due to
the $\Psi$ and $\Omega$ zero modes, to which are bound
a single $\Psi$ fermion with asymptotic four-momentum $p_{\mu}$  
and mass $m_{4}$ and a single $\Omega$ fermion with
asymptotic four-momentum $k_{\mu}$ and mass $\widetilde{m}_{4}$.

We next consider the operator equation \eqref{eq:phi-eq},
leaving aside issues of operator ordering, and
take the diagonal matrix element
between the $--$ soliton with $\Psi$ and $\Omega$ bound fermions of equal
mass, $\widetilde{m}_{4}=m_{4}$. The result is a partial differential
equation for the scalar form factor,
with a source term on the right-hand side obtained
from the following matrix element:
\beq
\label{eq:fermion-source-term-quantum}
\langle P^{\prime},-,-;\,  p, m_{4};\,  k, m_{4}|\,
f\,\left(\bar{\mathbf{\Psi}}\mathbf{\Psi}-\bar{\mathbf{\Omega}}\mathbf{\Omega}\right)
|P,-,-;\,  p, m_{4};\,  k, m_{4} \rangle\,,
\eeq
where $\mathbf{\Psi}$ and $\mathbf{\Omega}$ are quantum fields
(denoted by bold symbols).
For sufficiently low soliton and fermion momenta,
the leading contribution to the matrix element
\eqref{eq:fermion-source-term-quantum} is given by a
Fourier integral of the
right-hand side of \eqref{eq:fermion-source-term-classical}
in terms of spinor form factors $\psi_{l,r}$  and $\omega_{l,r}$,
which then cancel as shown by the further steps in
\eqref{eq:fermion-source-term-classical}.

Much more can and must be said on the quantum theory,
but that lies outside the scope of the present article,
which deals solely with solutions of the classical fields equations.

\section{Discussion}
\label{sec:Discussion}

The key input of the mechanism presented in Sec.~\ref{sec:Mechanism}
is the requirement that the spinor wave function be nonsingular and
localized in the fifth dimension (nonlocalized scattering-type solutions
exist but are not considered here).
More precisely, the wave functions of the left-handed and right-handed spinors have extra normalizable factors which are nonsingular functions of the fifth coordinate $w$.
This normalization condition together with the existence of the
fifth dimension bring about two important consequences: first, the bounded and discrete mass spectrum of the fermions in the four-dimensional Minkowski spacetime
[specifically, the mass spectrum is given by Eq.~\eqref{eq:integer-mass} for an integer ratio \eqref{eq:integer-F} of coupling constants
or by Eq.~\eqref{eq:mass-spectrum-noninteger-degree}
for a noninteger coupling-constant ratio larger than $3/2$]   
and, second, a hard-wired difference of left-handed and right-handed fermions [the difference being due to wave functions with
different associated Legendre functions \eqref{eq:vl-vr-solutions}
and \eqref{eq:vl-vr-solutions-noninteger-degree}].
Note that having the chirality of four-dimensional fermions distinguished by
their position in an extradimensional direction is precisely
what has been used to construct models of chiral lattice
fermions~\cite{Kaplan1992}.

It is also clear that the tangent hyperbolic function from the domain wall
plays a special role in the discussion of Sec.~\ref{sec:Mechanism}.
Recall that P\"{o}schl and Teller~\cite{PoeschlTeller1933} have already
studied a large class of sinus and hyperbolic-sinus potentials, whose corresponding Schr\"{o}dinger equations exhibit similar spectra. We may now ask the following question: is the tangent hyperbolic function absolutely necessary for obtaining a bounded and discrete spectrum of fermion masses?
We conjecture that
the answer is negative, based on following argument.

From the construction in Sec.~\ref{sec:Mechanism}, it is readily seen that, as long as we take the \textit{Ans\"{a}tze}
\eqref{eq:Psi-5D-separation-Ansatz} and \eqref{eq:4D-Ansatz}, and then require $v_{l}(w)$ and $v_{r}(w)$
to be localized functions in the fifth dimension,  
the spectrum would be necessarily discrete.
Some additional computations are needed
to demonstrate that the mass spectrum is bounded
for localized spinor solutions.
Replace $M\tanh(c\, w)$ in Eq.~\eqref{eq:firstordereqforv}
by a general function $\Phi(w)$. We then arrive at the following
coupled equations:
\bsubeqs \label{eq:first-order-general-phi}
\beqa
 -\partial_w v_{l} + m_{4}\,  v_{r}  &=& f\, \Phi \,  v_{l} \,,\\[2mm]
 +\partial_w v_{r} + m_{4}\,  v_{l}  &=& f\, \Phi \,  v_{r}  \,,
\eeqa\esubeqs
which give the following uncoupled equations:
\bsubeqs  \label{eq:second-order-general-phi}
\beqa
 -\partial^2_w v_{l} +\left(f^2\,\Phi^2-f\,\partial_w\Phi\right)\,v_{l}
 &=& m_{4}^2\, v_{l} \,,\\[2mm]
 -\partial^2_w v_{r} +\left(f^2\,\Phi^2+f\,\partial_w\Phi\right)\,v_{r}
 &=& m_{4}^2\,v_{r}  \,.
\eeqa\esubeqs
These last two equations give the eigenfunctions of two
Schr\"{o}dinger-type equations with potentials
proportional to \mbox{$(f^2\,\Phi^2  \mp f\,\partial_w\Phi)$}
and energy eigenvalue proportional to $m_{4}^2$\,.

In order to have localized wave functions from the
Schr\"{o}dinger-type equations \eqref{eq:second-order-general-phi},
there must be a deep enough potential-energy well.
If the scalar background field $\Phi(w)$ approaches different finite values at $\pm\infty$, the derivatives vanish asymptotically.
Then $\Phi^2$ may provide finite-height edges of a potential-energy well.
In order to obtain localized states, the energy of the ``particle''
(essentially $m_{4}^2$) cannot exceed the minimum height of the edges
of the potential-energy well.
Hence, the mass spectrum is bounded. However, due to the different signs in front of the terms $\partial_w \Phi$ in the Schr\"{o}dinger potentials
of \eqref{eq:second-order-general-phi}, the
mass  spectra for left-handed and right-handed spinors will, in general, not overlap. Without overlap, there would be no combined solutions to Eqs.~\eqref{eq:first-order-general-phi} and \eqref{eq:second-order-general-phi}. It is indeed a pleasant surprise
that the domain-wall solution of the scalar field equation
can produce two largely overlapping spectra. There are, in principle,
other potentials $\Phi(w)$ which can produce overlapping spectra and they
can be expected to give different numerical predictions for the fermion masses.
Further investigation of the pair of Schr\"{o}dinger-type equations
\eqref{eq:second-order-general-phi} is needed
(see also the discussion in
Ref.~\cite{RingevalPeterUzan2001}).

As a final remark, we note that the same type of analysis
applies to (1+1)-dimensional fermions moving
along a domain wall in 2+1 spacetime dimensions.
[Specifically, taking the three-dimensional
gamma matrices as $\Gamma^0=\sigma^1$, $\Gamma^1=i \sigma^2$, and
$\Gamma^2=i \sigma^3$, the calculation in $1+2$ dimensions directly
parallels the one in $1+4$ dimensions.]
The challenge for condensed matter physics is to provide a suitable domain wall (or an equivalent trapping mechanism)
and to tune the two Yukawa coupling constants to appropriate values.

\section*{\hspace*{-5mm}ACKNOWLEDGMENTS}
\noindent

The referee is thanked for useful comments.
This work has been supported, in part,
by the ``Helmholtz Alliance for Astroparticle Physics HAP,''
funded by the Initiative and Networking Fund of the Helmholtz Association.

\begin{appendix}

\section{Integrable solutions with noninteger Legendre degree}
\label{app:Integrable-solutions-with-noninteger-Legendre-degree}

It is clear from Ref.~\cite{MagnusOberhettingerSoni1966} that $P_{\nu}^{\mu}(x)$ and $Q_{\nu}^{\mu}(x)$
are two linearly independent solutions of the Legendre equation:
\begin{equation} \label{eq:GL}
(1-x^2)\,  \frac{d^2y}{dx^2}-2x\, \frac{dy}{dx} +\left[\nu(\nu+1)-\frac{\mu^{2}}{1-x^2}\right]\, y=0 \, ,
\end{equation}
where the degree $\nu$ and
the order $\mu$ are assumed to be real noninteger numbers.
In particular, $P_{\nu}^{\mu}(x)$ and $Q_{\nu}^{\mu}(x)$  are real
solutions on the real interval $(-1,1)$. By this restriction of the domain
(``on the cut''), the solutions are singularity free. The goal here is to show that there exist square-integrable solutions $y(x)$ to \eqref{eq:GL} with respect to the following measure and domain:
\beq
\int_{-1}^{1}\,\frac{dx}{1-x^2}\;|y|^2\,.
\eeq
Throughout this Appendix, square-integrability    
is always for this measure and domain,   
and for the noninteger parameters $\nu$ and $\mu$.
Recall from Sec.~\ref{sec:Mechanism} that the degree is given
by $\nu=F$ for the case of left-handed fermions
and by $\nu=F-1$ for the case of right-handed fermions,
while both cases correspond to the same order
$\mu=\pm \,(F^2 - m_f^2)^{1/2}$.

Since the solutions $P_{\nu}^{\mu}(x)$ and $Q_{\nu}^{\mu}(x)$
are singularity free on the real interval $(-1,1)$, their square-integrability depends solely on the behavior at the two ends of the interval [namely, $-1$ and $1$], because the measure diverges at these two ends. Without loss of generality,  $\nu$ is assumed to be a positive noninteger. We start with the solution $P_{\nu}^{\mu}(x)$
and discuss three results.

The first result is that
$\mu$ must be negative for $P_{\nu}^{\mu}(x)$ to be
square-integrable.   
This is because the behavior of $P_{\nu}^{\mu}(x)$ near $x=-1$ is problematic for $\mu>0$. Namely, the first equation on p.~197 in Ref.~\cite{MagnusOberhettingerSoni1966} says that, for $\mu>0$ and $x\sim -1^{+}$,
\begin{equation}\label{eq:first-eq}
P_{\nu}^{\mu}(x)\sim-2^{\,\mu/2}\,
\sin(\pi\nu)\, \pi^{-1}\, \Gamma(\mu)\, (1+x)^{-\mu/2}\, ,
\end{equation}
where $\Gamma(\mu)$ is the Euler Gamma function for argument $\mu$.
Since $\mu$ is positive, $\Gamma(\mu)$ is finite. Hence,
the numerical factor in front of $(1+x)^{-\mu/2}$ is finite.
Since the exponent of the $(1+x)$ term is negative
for $\mu>0$, $P_{\nu}^{\mu}(x)$ diverges at $x\sim-1^{+}$
(the superscript `$+$' here means ``just above'' $-1$)   
and is not  square-integrable on the chosen domain. Thus, the only possibility is $\mu <0$ (recall that $\mu$ has been assumed to be noninteger in this Appendix).

The second result is that
$\nu+\mu$ must be a non-negative integer for $P_{\nu}^{\mu}(x)$ to be  square-integrable.
Again, it is the behavior of $P_{\nu}^{\mu}(x)$ near $x=-1$ that causes problems for the case $\nu+\mu<0$. The second equation on p.~197 in Ref.~\cite{MagnusOberhettingerSoni1966} says that, for $\mu<0$ and $x\sim-1^{+}$,
\begin{equation}\label{eq:second-eq}
P_{\nu}^{\mu}(x)\sim 2^{-\mu/2}\,
\frac{\Gamma(-\mu)}{\Gamma(1+\nu-\mu)\,\Gamma(-\nu-\mu)}\, (1+x)^{\,\mu/2}\, .
\end{equation}
Now $\Gamma(-\nu-\mu)$ is finite for $\nu+\mu<0$.
Since $\mu$ is negative, $\Gamma(-\mu)$ is also finite. Hence, all numerical factors in front of $(1+x)^{\,\mu/2}$ are finite. For $\mu<0$, $(1+x)^{\,\mu/2}$ goes to infinity as $x$ approaches $-1$.
Then $P_{\nu}^{\mu}(x)$ cannot be  square-integrable on the domain for the
case considered. The only possibility left is for the other case, with $\mu<0$ and $\nu+\mu\geq 0$.

If $\nu+\mu$ is greater than or equal to $0$ but not an integer, $\Gamma(-\nu-\mu)$ is still finite and the other numerical factors
in Eq.~\eqref{eq:second-eq}, too. Thus $P_{\nu}^{\mu}(x)$ cannot be  square-integrable on the domain. However, if $\nu+\mu$ is a non-negative integer, $\Gamma(-\nu-\mu)$ becomes infinite and independent from $x$. Since $\Gamma(-\nu-\mu)$ appears in the denominator in the above equation, it may cancel the divergence from $(1+x)^{\,\mu/2}$. Thus, it is possible that
$P_{\nu}^{\mu}(x)$ becomes  square-integrable on the domain,
which is, in fact, to be discussed as the next result.

The third result is that
$P_{\nu}^{\mu}(x)$ is  square-integrable
if $\mu$ is less than $0$  and $\nu+\mu$ is a non-negative integer.
If $\nu+\mu$ is an integer, $P_{\nu}^{\mu}(x)$ is either odd or even on the domain $(-1,1)$, as implied by the 7th equation on p.~170 of Ref.~\cite{MagnusOberhettingerSoni1966}:
\begin{eqnarray}
P_{\nu}^{\mu}(-\widehat{x})& = & P_{\nu}^{\mu}(\widehat{x})\,\cos [\pi(\nu+\mu)] -2\pi^{-1} Q_{\nu}^{\mu}(\widehat{x})\,\sin[\pi(\nu +\mu)]
\nonumber\\
                 & = & \pm P_{\nu}^{\mu}(\widehat{x})\, ,
\end{eqnarray}
for $\widehat{x}\in(0,1)$.
In addition, the second equation on p.~192 in Ref.~\cite{MagnusOberhettingerSoni1966} reads, for the case $\mu<0$ and $\nu+\mu\in \{1,2,3 ...\}$,
\begin{equation}
\int_{0}^{1}(1-x^2)^{-1}\, \left[P_{\nu}^{\mu}(x)\right]^{2} dx =
-\frac{1}{2}\, \mu^{-1}\, \frac{\Gamma(1+\nu+\mu)}{\Gamma(1+\nu-\mu)}\, .
\end{equation}
According to Sec.~3.12 of Ref.~\cite{Erdelyi1953}, the above integral is originally due to Barnes.    
Combining the last two equations, one obtains
\begin{equation}\label{eq:Barnes-integral}
\int_{-1}^{1}(1-x^2)^{-1}\, \left[P_{\nu}^{\mu}(x)\right]^{2} dx =-\mu^{-1}\, \frac{\Gamma(1+\nu+\mu)}{\Gamma(1+\nu-\mu)}\, .
\end{equation}
Note that the right-hand side of the last equation is finite, making $P_{\nu}^{\mu}(x)$  square-integrable for
$\mu<0$ and $\nu+\mu\in \{1,2,3 ...\}$.
Remark also that the structure of the right-hand side of
Eq.~\eqref{eq:Barnes-integral} corresponds to
that of Eq.~\eqref{eq:Plm-norm}.

If $\nu+\mu=0$, we can use Eq.~(8.6.17)
from Ref.~\cite{AbramowitzStegun1965}, which reads
\begin{equation}
P_{\nu}^{-\nu}(\cos\theta)=\frac{(\sin \theta)^{\nu}}{2^{\nu}\,\Gamma(\nu+1)}\,,
\end{equation}
where $\cos\theta$ is a parametrization of $x$, with $\theta\in(0,\pi)$.
It is now clear that $P_{\nu}^{-\nu}$ is  square-integrable
if and only if $\nu>1/2$.   
The condition $\nu>1/2$ translates
into $F>1/2$ for left-handed fermions and
into $F>3/2$ for right-handed fermions.   

To sum up, for the case that both $\mu$ and $\nu$ are noninteger,
the necessary and sufficient condition for $P_{\nu}^{\mu}$ to be  square-integrable, with respect to the chosen measure and domain, is that $\mu$ is negative, $\nu$ is larger than $1/2$,  
and $\mu+\nu$ is a non-negative integer.
In short, $P_{\nu}^{\mu}(x)$ is  square-integrable for a
given degree $\nu$ with
\bsubeqs\label{eq:nu-mu-ranges}
\beqa
\nu &\in& \mathbb{R}\backslash  \mathbb{Z}\,,
 \\[2mm]
\nu &>&  1/2 \,,   
\eeqa
only if the order $\mu$ is given by
\beqa
\mu &\in& \{
-\nu,
-\nu+1, -\nu+2, ..., -\nu+\floor{\nu} \}  \,,
\eeqa
\esubeqs
where $\floor{\nu}$ denotes the largest integer that is
not greater than $\nu$
(in the mathematics literature, $\floor{x}$
is called the floor or \textit{Entier} function of the real number $x$).

It is also necessary to check if $Q_{\nu}^{\mu}(x)$ in certain cases is  square-integrable with respect to the chosen measure. The check is done in a similar way as for $P_{\nu}^{\mu}(x)$. The answer is affirmative, only if both $\nu$ and $\mu$ are positive half odd integers and $\mu<\nu$.
For these cases, the fourth equation on p.~170 in Ref.~\cite{MagnusOberhettingerSoni1966} says
\begin{eqnarray}
Q_{\nu}^{-\mu}(x)& = & \frac{\Gamma(\nu-\mu+1)}{\Gamma(\nu+\mu+1)}\,
\left[Q_{\nu}^{\mu}(x)\,\cos(\pi\mu) +\frac{\pi}{2} \, P_{\nu}^{\mu}(x)\,\sin(\pi\mu)\right]
\nonumber\\[2mm]
& = & \pm \frac{\pi}{2} \,\frac{\Gamma(\nu-\mu+1)}{\Gamma(\nu+\mu+1)}\, P_{\nu}^{\mu}(x)\, ,
\end{eqnarray}
and the conclusion is
that $P_{\nu}^{-\mu}(x)$ and  $Q_{\nu}^{\mu}(x)$ give essentially the same solution of \eqref{eq:GL}. Thus, these cases are already included in the previous discussion, up to a change of sign for $\mu$.

It should be pointed out that linear combinations of $P_{\nu}^{\mu}(x)$ and $Q_{\nu}^{\mu}(x)$ have not been checked. We assume that this gives no additional interesting cases.

Considering the two general Legendre equations for $v_r(w)$ and $v_l(w)$,
we obtain the fermion mass spectrum
as given by Eq.~\eqref{eq:mass-spectrum-noninteger-degree} in the main text.

\section{Massless spinor solution}
\label{app:Massless-spinor-solution}

Another type of spinor solution holds for the
massless case, $m_f=0$ in Eqs.~\eqref{eq:partial-vl-vr-eqs}.
Taking a positive coupling-constant ratio, $F \equiv f/\lambda > 0$,
it is readily verified that a solution is given by
\bsubeqs\label{eq:vl-vr-solutions-massless}
\beqa
\label{eq:vl-solution-massless}
v_{l}(s) &\propto&  (1-s^2)^{\,F/2}\,,
\\[2mm]
\label{eq:vr-solution-massless}
v_{r}(s) &=& 0\,,
\eeqa\esubeqs
where $v_{l}(s)$ is  square-integrable.
A nonzero solution for $v_{r}(s)$ does
exist, proportional to $(1-s^2)^{-F/2}$,
but is not  square-integrable.
The solution \eqref{eq:vl-vr-solutions-massless}
corresponds to a chiral spinor and is well known in the literature
(cf. Refs.~\cite{Rubakov2001,Kaplan1992,JackiwRebbi1976}).

Strictly speaking,
the solution \eqref{eq:vr-solution-massless}
is not localized on the domain wall at $s=0$.
For this reason, we have not included
the mass value $0$ in the discussion of
Sec.~\ref{subsec:Spinor-solutions-and-mass-fermion-spectrum},
but massless chiral spinors are definitely part
of the low-energy physics.

\end{appendix}


\end{document}